# Combining Keystroke Dynamics and Face Recognition for User Verification

Abhinav Gupta, Agrim Khanna, Anmol Jagetia, Devansh Sharma, Sanchit Alekh, Vaibhav Choudhary
Department of Information Technology
Indian Institute of Information Technology
Allahabad, India
{abhinavgupta2004, khannaagrim, anmoljagetia, devanshs212, sanchit.alekh, choudharyvaibhav132}@gmail.com

*Abstract*—The massive explosion and ubiquity of computing devices and the outreach of the web have been the most defining events of the century so far. As more and more people gain access to the internet, traditional know-something and have-something authentication methods such as PINs and passwords are proving to be insufficient for prohibiting unauthorized access to increasingly personal data on the web. Therefore, the need of the hour is a user-verification system that is not only more reliable and secure, but also unobtrusive and minimalistic. Keystroke Dynamics is a novel Biometric Technique; it is not only unobtrusive, but also transparent and inexpensive. The fusion of keystroke dynamics and Face Recognition engenders the most desirable characteristics of a verification system. Our implementation uses Hidden Markov Models (HMM) for modeling the Keystroke Dynamics, with the help of two widely used Feature Vectors: Keypress Latency and Keypress Duration. On the other hand, Face Recognition makes use of the traditional Eigenfaces approach. The results show that the system has a high precision, with a False Acceptance Rate of 5.4% and a False Rejection Rate of 9.2%. Moreover, it is also future-proof, as the hardware requirements, i.e. camera and keyboard (physical or on-screen), have become an indispensable part of modern computing.

## I. INTRODUCTION

Over the years, a lot of effort and time has been put into improving the security and privacy of user accounts. Yet, throughout history, we have experienced numerous occasions of intrusion, impersonation and breach of confidentiality. Various static and well as continuous verification systems have been implemented using a plethora of different approaches, such as fingerprint recognition, face recognition, iris scan, traditional PIN and password, modern approaches to passwords such as dynamic one-time password generators, picture-based and graphical passwords, cognitive passwords and envaulting technology. Biometric Verification essentially refers to the process of validation by means of biometric traits that are unique to the user. Biometric traits include fingerprint, face, iris, signature and voice. They offer several advantages over PIN and password method, the most important one being that they are quite difficult, if not impossible to impersonate. Biometric methods also eliminate the need to remember specific passwords, and do not need to be changed periodically. However, certain drawbacks do exist while using them, which include inefficiency of the hardware to get good enough samples or the inefficiency of the algorithm to recognize patterns under certain conditions. The tradeoff between the benefits and the drawbacks can be evaluated depending on the cost, security and other constraints of the application.

Many of these biometric verification approaches are either expensive to use or hardware intensive, which makes them unsuitable for such a wide range of users. Also, each and every of the above mentioned techniques have some inherent shortcomings, and are vulnerable to attacks and hacks. Therefore, one of the approaches can be the usage of a combination of modalities instead of a single modality verification technique. The idea behind such an approach is to cover up the deficiencies of one biometric modality with the other, which might lead to a reduction in certain breaches for each of the modalities, as well as offer a much better performance.

Moreover, most of the current user security systems, once the user identity has been verified at login, system resources are available to the user until the user exits the system. This may be useful for low-security environments, but can lead to session hijacking in which the attacker targets a post-authenticated session [4]. Moreover, recent research findings have shown that PINs and passwords are no longer secure, owing to the ever increasing computer users [1]. Continuous Authentication is a subject of importance due to the fact that a logged station is still vulnerable to intrusion and unauthorized access, and behavioral metrics such as keystroke dynamics are very helpful in this kind of a scenario.

Our idea is to build a cost-effective as well as efficient user-identification system which would be unobtrusive to the user as well as shelves off some of the privacy concerns in and around the web, such as social networks, online examinations and likewise. An amalgamation of keystroke biometrics and face recognition provides an excellent solution to the problem of precise user-verification. The modalities were chosen on the basis of their user-friendliness, unobtrusiveness and the fact that they did not require any additional hardware [3].

Keystroke dynamics are an effective behavioral biometric, which captures the habitual patterns or rhythms an individual exhibits while typing on a keyboard input device. These rhythms and patterns of tapping are idiosyncratic, in the same way as handwritings or signatures, due to their similar governing neurophysiological mechanisms [2]. With our implementation we were able to create a user verification software that relied on both the face and keystroke recognition. By overcoming the drawback of one modality with another, the implementation was able to achieve an accuracy level that gives sufficient confidence in the system for practical real-life execution and scope for further work and improvement on the subject.

## II. RELATED WORK

The individuality of a person's keystroke pattern was intially reported by Joyce and Gupta [5]. Since then, a lot of work has been done on Keystroke Biometrics. Several classifiers such as Statistical Methods, Distance Measures, K-nearest neighbor approach, several Neural Network architectures such as Perceptron, Probabilistic Neural Networks, Weightless Neural Networks, Back-propagation, Adaptive Resonance Theory techniques and Support Vector Machines have been used in the past. Different kinds of classification techniques have been employed for static as well as dynamic text. A multi-modal user-verification system was developed by Filho and Friere [6] by fusing Keystroke Patterns with Voice Recognition, but was limited to 10 users only.

Some researchers have experimented fusing different algorithms of the same modality. For instance, in [7], a few different keystroke dynamics implementations were fused with an improvement of the Error Equal Rate (EER), even if lesser than 40 users had used the system. In [8], two keystroke dynamics systems were fused together by using weighted sums for 50 users. However, not much information on the computation of weight is provided. Another method that has been widely used to improve the efficiency is by using fusion techniques within the different modalities. In [9], multi-modality has been achieved by combining features such as fingerprints, speech and facial images. Support Vector Machines have yielded pretty good results too [10], especially when using user-specific classifiers. Yu and Cho [11] sought to improve the performance of keystroke dynamics by using a three-step approach. Several attempts have also been made to use different types of algorithms such as Genetic Algorithms and an ensemble based on feature selection by Yu and Cho [11], Nearest Neighbor Approach and Multilayer Perceptron by Killourhy & Maxion [22].

Previous work on this subject has not focused enough on combining two dissimilar biometric techniques such as keystroke biometrics and face recognition. One such work by Giot et al. [25] attempted to fuse statistical techniques, support vector machines and rhythmic typing patterns for keystroke data, and eigenfaces and double key-point association for face recognition. They were able to show that a combination of the biometric modalities gives a better precision than any single one evaluated individually. By using Hidden Markov Models for modeling keystroke patterns, our work attempts to implement a pattern recognition algorithm which is more suited to capture temporal data than support vector machines.

## III. DATA COLLECTION AND PROCESSING

### A. Collection and Processing of Keystroke Data

The keystroke pattern of a user can be uniquely identified by the relation between the key press times and the key release times. Gaines, Lisowski, Press and Shapiro [12] showed that in most of the implementations, measures such as keystroke latencies and keystroke duration have proven to be an excellent indicators of the user patterns. For the collection of keystroke data, a Java program is used which records the timing of each keypress with a precision of 1 ms. With this data, we create the two useful feature vectors: keypress duration and keypress latency.

$$keyduration_t = releasetime_t - presstime_t \quad (1)$$

$$keylatency_t = presstime_t - releasetime_{t-1} \quad (2)$$

where $keyduration_t$ is the time duration for which the $t_{th}$ key was pressed, $releasetime_t$ is the time at which the $t_{th}$ key was released, $presstime_t$ is the time at which the $t_{th}$ key was pressed and $keylatency_t$ is the time duration between the $t_{th}$ key release and the $t+1_{th}$ key press. The collected data is normalized to make it independent of the typing speeds that may vary depending on the condition of the user.

$$normalized\_duration_t = \frac{keyduration_t}{releasetime_T - presstime_0} \quad (3.1)$$

$$normalized\_latency_t = \frac{keylatency_t}{releasetime_T - presstime_0} \quad (3.2)$$

where $normalized\_duration_t$ is the normalized time duration of the $t_{th}$ key and $normalized\_latency_t$ is the normalized time latency of the $t_{th}$ key. T refers to the final key press-release iteration. Figure 1 is a schematic diagram that shows the presstime, releasetime, keyduration and keylatency with respect to for the user input string "GOOSEBERRY". The diagram highlights the fact that keylatency latency can be a negative value as well.

This resulting data is now stored in an encrypted user-specific file, i.e, a separate file for each user. Each observation obtained is stored as a vector of keypress duration and latencies, and in turn, the vectors are separated by semicolons to make the parsing convenient. For every new iteration, a line break is added.

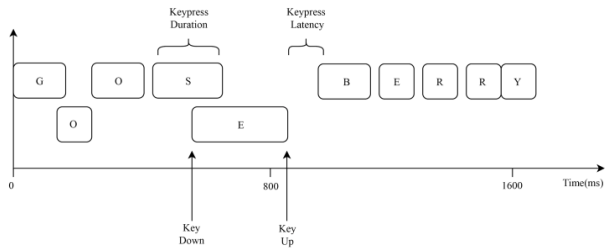

Figure 1. Collection of Keystroke Latency and Duration data from input string 'GOOSEBERRY' *(not drawn to scale)*

### B. Collection and Processing of Facial Data

The facial data of the users is collected using an external camera or an integrated webcam. For a new user, the software picks up twenty captures of the facial data, and uses a combination of Principal Component Analysis (PCA) and the Linear Discriminant Analysis (LDA) algorithms. For reducing the dimensionality, the pictures are converted into grayscale, and a code written in Python identifies the facial part of the image and crops the image in such a way that the face is magnified. Using the training dataset so created, the Eigenfaces and Fisherfaces are generated.

## IV. PATTERN RECOGNITION USING THE HIDDEN MARKOV MODEL

A Hidden Markov Model (HMM) is a statistical model where the output or the emission is observable but the states through which the model went through in order to project the output, is unknown or hidden and the analysis of the model seeks to obtain this state sequence. The primary reason for using a HMM for modeling the keystroke verification process is that typing patterns are essentially temporal patterns with stochastic nature and that the HMM is known to be quite efficient for modeling events that are distributed over temporal range without a deterministic value or position.

Our HMM has 2 states, which don't have a physical significance but may be thought of as a state where the user knows the password and is confident while typing and the other being the state where the user might have doubts about the password, i.e, what key has to be pressed next and that gets reflected in his typing style.

A multivariate gaussian probability distribution is used to model the observation sequences, where the keypress duration and keypress latency are the two random variables and it is assumed that the two random variables have a correlation between them. After that with Baum-Welsh algorithm [13], the unknown parameters of the HMM, i.e. the transition and emission probabilities are estimated. Baum-Welsh algorithm finds the unknown parameters of the HMM by determining the maximum likelihood of the parameters given the observed feature vector. Baum-Welsh algorithm itself uses the Forward-Backward Algorithm that estimates, for all the states, the posterior probabilities given the sequence of emissions. The first pass goes forward in time and the second pass goes backward in time and hence the name 'forward-backward'.

After applying the Baum-Welsh algorithm to find the unknown parameters of the HMM, an equilibrium point is achieved. Figure 2 represents an equilibrium condition for a user that we obtained during training the HMM model for that user, here the values mentioned along with the states are the mean values of the feature vector in that particular state. Since no tractable solution can exactly find the unknown parameters, the values represented by the arrows are the state transition values that were achieved after 20 iterations of Baum-Welsh algorithm. Finally the Hidden Markov Model defined by the states, the initial probabilities of those states, the transition and emission probabilities are stored in another encrypted file. Each user has her own trained Hidden Markov Model, stored separately.

### A. The Generalization Process

The Each time a user's identity has to be verified, the observation sequence in form of keypress latency and keypress duration, is recorded and stored in another file. Then the pattern is scored with the trained Hidden Markov Model, that is first retrieved and decrypted. The scoring is done with the help of Viterbi Algorithm[14] that outputs the Viterbi Path. Viterbi algorithm is a dynamic algorithm that is used to find the most likely sequence of the states that the model went through to give out the observed output or emission. And this sequence of states that Viterbi algorithm results in, is called as the Viterbi Path. Next, the probability that the observation sequence is legitimate, given the trained HMM of the user and the state sequence computed by the Viterbi Algorithm, is obtained.

### B. Figures and Tables

For calculating the threshold, the previous typing samples of each of the users are scored against the trained Hidden Markov Model of the user and the mean and standard deviation of the probability values is extracted. This provides the information of how the probability that the observation sequence is legitimate, given the trained hmm of the user and the state sequence computed by the Viterbi algorithm, varies with the change in typing pattern of the user that is dependent on the physical conditions.

After that the threshold is defined by the standard deviation around the mean of the probabilities. The threshold can then be adjusted by either increasing or decreasing the acceptance region around the neighborhood of the mean probability, depending on the precision required by the application. For our implementation, we simply used an acceptance threshold given by mean plus standard deviation. A trade-off exists in increasing or decreasing the threshold, and a desired value can be set on the basis of the severity of consequences that may follow a false acceptance.

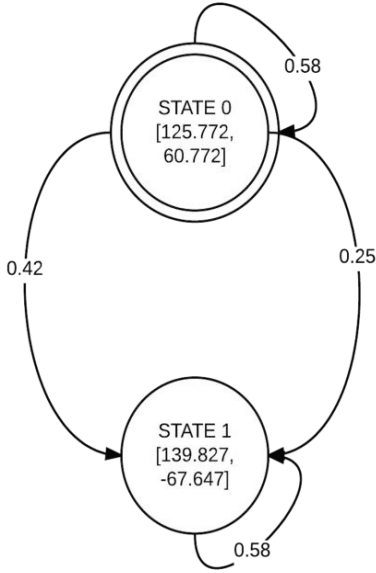

Figure 2. State transitions of a user's trained HMM. The arrows represent the transition probabilities of the HMM and the values along with the states represent the mean values of the feature vector.

## V. Face Recognition using Traditional Eigenfaces Approach

The digital interpolation of a face is a complex and intricate representation, and therefore requires carefully crafted techniques for analysis. For this reason, Face Recognition has been an active area of research for a long time. In most of the cases, Face Recognition involves a dimensionality reduction task, since the computational tools available right now are not suited to operate in such a high-dimensional space. PCA is one such technique. Also called Eigenspace Projection or Karhunen-Loeve Transformation, PCA relies on finding the eigenvectors and eigenvalues of the covariance matrix. LDA, also called as Fisher's Discriminant Analysis is applied after initial feature extraction. LDA tries to maximize the between-class scatter and minimize the within-class scatter matrices. It does so by calculating the eigenvectors and eigenvalues of the scatter matrices.

After generating the Eigenfaces and Fisherfaces from the transformed sample data, the Face Recognition module is ready. When a user attempts log in, his image data is collected via the external camera/webcam and transformed. Thereafter, a simple Euclidean Distance is calculated from the Eigenface and projected Fisherface. The image is classified into the class with the least Euclidean distance from the resulting projected Fisherface. Figure 3 illustrates a sample the sample Eigenfaces and Fisherface generated using the test input data.

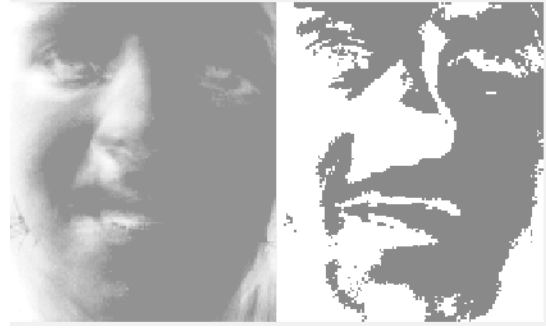

Figure 3. Eigenface and Fisherface obtained from a sample input image data

## VI. Combining the Two Modalities

After The final decision on whether to grant or deny access to a user is based on the combined score of the two modalities. For our implementation, different integrators i.e. sum, min, max and product, were used to combine the results. The score of the user being himself given the keystroke biometrics claims so, and the score of him being an imposter given that keystroke biometric claims so, is calculated. These probabilities are then operated upon by the above mentioned score-based integrators to give the score values.

Depending on the security requirements of the application and the consequences of an unauthorized access, more complex integrators might be possible that are more sensitive and result in a positive decision only if the observations of both the modalities are very close to the expected value. Similarly for the data received from the face recognition module, we obtain the score that the user is himself or an imposter given the face biometric data claims so.

Based on observations from several tests with the different integrators, we came to the conclusion that the product integrator was the most suited for our application, as it led to lower error rates. Equations (4) and (5) depict the usage of the product integrator in our model.

$$S_{usertrue} = P(user_{true} | key_{usertrue}) \times P(user_{true} | face_{usertrue}) \quad (4)$$

$$S_{userfalse} = P(user_{false} | key_{userfalse}) \times P(user_{false} | face_{userfalse}) \quad (5)$$

where $S_{usertrue}$ and $S_{userfalse}$ are the scores for a user being genuine and an imposter respectively, $user_{true}$ and $user_{false}$ are the instances when the user is true and false respectively, $key_{usertrue}$ and $key_{userfalse}$ are the instances when keystroke data says user is true and false respectively, and $face_{usertrue}$ and $face_{userfalse}$ are the number of instances when facial data says the user is true and false respectively. Finally, a decision is made on the basis of the probability values hence calculated. The user is verified if the probability of him being genuine exceeds the probability of him being an imposter, i.e. when $S_{usertrue} > S_{userfalse}$ and vice versa.

## VII. RESULTS AND DISCUSSION

In previous works on the subject, most of the authors have reported their performance in terms of error metrics False Acceptance Rates (FAR) and False Rejection Rates (FRR). FAR measures the number of false matches as a proportion of total number of impostor match attempts, while FRR, on the other hand, measures the number of false rejections as a proportion of the total number of genuine match attempts.

$$\text{FAR} = \frac{No. \text{ of false matches}}{\text{Total no of imposter attempts}} \quad (6)$$

$$\text{FRR} = \frac{No. \text{ of false rejections}}{\text{Total no. of genuine match attempts}} \quad (7)$$

A high FRR is preferred for systems where the security is desired to be infallible [21], whereas for systems where no such stringent security is required, a high FAR is preferred. Table I illustrates the various approaches used by various researches for modeling keystroke dynamics and the error rates reported by them.

TABLE I. KEYSTROKE DYNAMICS MODELING ARCHITECTURES

| S. No. | *Study* | *Technique Used* | *FAR (%)* | *FRR (%)* |
|---|---|---|---|---|
| 1 | Joyce and Gupta [5] | Statistical | 0.25 | 16.36 |
| 2 | Monrose and Rubin [3] | Statistical | 7.9 (Combined) | |
| 3 | Gunetti and Picardi [16] | Neural Network | 0.005 | 5 |
| 4 | Sheng et al.[17] | Parallel Decision Trees | 0.88 | 9.62 |
| 5 | Leggett and Williams [18] | Digraph Test | 5.5 | 5 |
| 6 | Bleha et al. [19] | Min. Distance Classifier | 2.8 | 8.1 |
| 7 | Hempstalk et al. [20] | One-class Gaussian | 11.3 | 20.4 |

By combining an implementation of Keystroke Dynamics using HMM and Face Recognition, we were able to generate an FAR of 5.4% and an FRR of 9.2%. We took a sample of 500 inputs from a user using a Java program that recorded the timing of each keypress with a precision of 1 ms. On an average, 27 inputs were falsely accepted, whereas 46 inputs were falsely rejected. The biggest advantage that HMMs offer, is that it is extremely intuitive to process vector inputs. Therefore, by adding features such as Relative Key Event Order and Average Typing Speed, the accuracy of the system can be further improved.

During our experiments with the verification system, a very interesting observation that we gathered, was that the password pattern can be made unique and more secure by varying the typing speed and a strong alphanumeric password, whereas a constant typing speed along with a simple string can be relatively easy to imitate. This is analogous to the practice in which users are encouraged to use a strong alphanumeric combination as an account password. We found out, that in cases where the user intentionally used a certain pattern while typing, the FAR came down by 26%.

Both Keystroke Biometrics and Face Recognition pose some inherent challenges. Keystroke Patterns are heavily dependent on posture, device, keyboard layout etc. The emotional state of the user also plays a determining factor on the typing speed. Khanna and Sasikumar [23] found that a negative mood state led to a 70% reduction in typing speed compared to an 83% increase in the typing speed in a positive state. Emotions also have a significant effect on typing, as reported by Epp et al. [24]. Similarly, the surface on which the computer is placed, type and brand of computer used could also have an impact on the efficiency with which a person can be accurately classified. In the same way, Face Recognition is highly sensitive to the lighting conditions, and give poor results in low-light environment. Similarly, physical changes over a period of time such as growing a beard, wearing spectacles etc. can lead to misclassification.

## VIII. CONCLUSION

In this paper an efficient way to verify user identity has been introduced and we conclude that the combination of keystroke biometric and face recognition is an effective way to implement a secure user-verification system. Without the need to analyze many features, especially for the keystroke biometric and a relatively simple classification technique, the fusion of keystroke and face recognition has proved to be a very promising approach. Another primary advantage of our system is the feasibility of the implementation as it only requires a software installation to work on any workstation.

There is immense scope for future work, one of the possible direction could be to represent the input data using a different Hidden Markov Model. The largest strength of the Hidden Markov Model in the case of Keystroke Dynamics is that it can build an efficient model for vector data sets. That makes it relevant to expand the keystroke verification by adding more feature vectors and using more metrics, such as Relative Key Press Order and Average Typing Speed. The implementation of our algorithms would work if the text is kept known and constant, such as a password. However, for user authentication in case of free text data, this approach would fail. In fact the ordinary neural network which receives a fixed number of features as an input vector cannot be directly used in this case [15]. In this situation, a different approach based on the use of Self-Organizing Maps are used.

Another direction for future work could towards improvising the classifier that combines the decision of the two modalities. In this paper, we fused two different methods with high efficiency to make keystroke biometrics less prone to forgery, and to make it usable for granting access in critical systems. We would also like to explore the possibility of using techniques such as Hopfield Networks or unsupervised learning techniques to make the module more robust. In the future, we would also like to create a data bank of a large number of users so that the more study can be done on the pattern of keystroke data for masses. Since the intruder is more likely to be interacting with a GUI in almost every case, mouse and trackpad behaviors and movements, along with keystrokes and facial data may be very helpful for further increasing the efficiency of decision making. We intend to extend our work to some of those aspects as well.

ACKNOWLEDGMENT

We would like to heartily acknowledge the guidance, motivation and support of Prof. Sudip Sanyal (Professor, Indian Institute of Information Technology Allahabad). The discussions our team had with him went a long way in shaping the outcome of our project. We are indebted to his contribution.

We would also like to place on record, our sincere thanks to Dr. Rahul Kala for his invaluable feedback and exemplary guidance, and also for proof-reading our work.